\documentclass[epj]{svjour}          

\usepackage[T1]{fontenc}

\smartqed  

\usepackage{graphicx}
\usepackage{mathptmx}      
\usepackage{flushend}
\usepackage{cite}

\usepackage{changes}
\usepackage[colorlinks,citecolor=blue,urlcolor=blue,linkcolor=blue]{hyperref}
\usepackage{subfigure}
\usepackage{amssymb}
\usepackage{multirow}
\usepackage{amsmath}
\usepackage{amsfonts}
\usepackage{amssymb}
\usepackage{subfigure}
\usepackage[utf8]{inputenc}
\usepackage{slashed}
\hypersetup{pdfstartview=}

\begin{document}

\title{Single top production via FCNC couplings at an $e^{-}e^{+}$ collider with center-of-mass energy $\sqrt{s}=240$ GeV}



\author{M. A. Arroyo-Ure\~na\inst{1}    \and
        J. Lorenzo D\'iaz-Cruz\inst{2} \and
        R. Gait\'an\inst{1} \and
        J.H.  Montes  de  Oca\inst{1} \and
        T. A. Valencia-P\'erez\inst{3}
}


\institute{
Departamento de F\'isica, FES Cuautitl\'an,\\
Universidad Nacional Aut\'onoma de M\'exico\\
Estado de M\'exico, 54770, M\'exico\and
Facultad de Ciencias F\'isico - Matem\'aticas,\\
  Benem\'erita Universidad Aut\'onoma de Puebla,\\
 C.P. 72570, Puebla, Pue., Mexico\and
 Instituto de F\'isica\\
 Universidad Nacional Aut\'onoma de M\'exico, C.P. 01000, CDMX, M\'exico.
}

\date{Received: date / Revised version: date}




\abstract{
	Aspects of top quark physics can be studied at a future Circular Electron-Positron Collider (CEPC). In particular, Flavor-Changing Neutral Current interactions (FCNC) of the top quark can be studied even with a center-of-mass energy from $\sqrt{s}\gtrsim m_t$ to $\sqrt{s}=240$ GeV, through the production a single top quark in association with a charm quark, which can arise in several scenarios of new physics. We explore here this signal within the Left-Right Mirror Model, which is a Left-Right model that includes mirror fermions; this model is motivated as a possible solution to the strong CP problem and by its rich phenomenology. We find that for particular scenarios of the model parameters, which are in agreement with the most up-to-date experimental constraints, the signal coming from the process $e^+e^-\to V \to tc$ ($V=Z,\,Z^{\prime}$) provide a detectable signal and thus it can give evidence for FCNC interactions; in these scenarios the mass of a new neutral gauge boson can be as high as $m_{Z'}=7$ TeV. 
} 

\maketitle






\section{Introduction \label{Introduction}}

The top quark is the heaviest fundamental fermion within the Standard Model (SM), and its large mass ($m_{t}$=172.76 GeV) is associated with a large  Yukawa coupling, which is of $O(1)$ and quite different from the other SM fermions. This aspect of the flavor problem  has been studied for some time, and it is believed that its resolution will require physics beyond the SM (BSM) \cite{Agashe:2005vg}.
Many of these models follow different approaches (including new symmetries or extra particles) and predict new phenomena, such as Flavor-Changing Neutral Currents (FCNC), which could be used to probe the assumptions involved in the construction of those models \cite{Fuchs:2020uoc}.
Currently, FCNC have been observed only for $d$-type quarks, in particular for the $b-s$ and $d-s$  transitions (e.g. $b\to s\gamma$, $B_s \to \mu^+  \mu^-$). However, the corresponding  FCNC process for the up-type quarks have not been observed,
neither the related Lepton Flavor Violating (LFV) transitions for leptons, or for SM particles, such as the Higgs or the $Z$ bosons \cite{Heinemann:2019trx}. 

The top quark  FCNC decays  include the modes:  $t\to cX$, with $X= h, Z, \gamma, g$, which were calculated first in \cite{DiazCruz:1989ub,Eilam:1990zc}, and later on they were studied for other models \cite{DiazCruz:2001gf,  AguilarSaavedra:2004wm, Arroyo:2013kaa, Bolanos:2019dso, Botella:2015hoa, Li:1993mg, Eilam:2001dh, Yang:2013lpa, Gaitan:2017tka} including its search at the LHC \cite{Arroyo-Urena:2019qhl, Diaz-Cruz:2019emo, Chen:2013qta}. One can also consider the FCNC decays of heavy particles ($H^{\prime},Z^{\prime}$) that can decay into $t\bar{c}$ \cite{Altunkaynak:2015twa,Arroyo-Urena:2019fyd,Li:2021qyo}. Future colliders offer the possibility to probe models of new physics including, among other goals, searches for FCNC processes \cite{Apollinari:2017cqg, Benedikt:2018ofy, Arkani-Hamed:2015vfh}. 
 
It is possible to search for top FCNC decays with pair production \cite{Shi:2019epw}.
The Circular Electron Positron Collider (CEPC) could produce about one trillion $Z$ gauge bosons, 100 million $W$ gauge bosons and one million Higgs bosons over the course of 7 years. A large amount of bottom quarks, $\tau$-leptons produced by decays of the $Z$ boson makes also the CEPC a $\tau$-factory. However, the production of top quark pairs requires at least a center-of-mass energy of about $2m_{\text{t}}\approx 345$ GeV, beyond the currently planned CEPC energy \cite{CEPCStudyGroup:2018ghi}. 
But, could CEPC considered a source of top quarks?  Namely, it is an  interesting question to ask whether one can study top FCNC couplings at the CEPC, operating below the production threshold \cite{Shi:2019epw}.

In this paper, we study this possibility using the production of a single top quark in association with a charm quark which proceeds through top FCNC interactions, for a particular realistic scenario within an extension of the SM. Namely, we shall work within the context of the so-called  Left-Right Mirror Model (LRMM), namely is a Left-Right (LR) model which includes mirror fermions that are singlets under the $SU(2)_L$ of the SM, but transform as doublets under a new  $SU(2)_R$ gauge symmetry. This model was motivated as a possible solution to the strong CP problem; the symmetry that forbids the strong CP-violating term is precisely the parity symmetry ($P$) incorporated into the model \cite{Barr:1991qx}. 
 We shall obtain a region of parameters of the model that defines a realistic scenario, in the sense that it satisfies all available experimental constraints. This model has an extended spectrum with several new particles, but here we will
concentrate in the SM particle content, with new FCNC couplings, supplemented with a heavy neutral gauge boson ($Z'$).

This letter is organized as follows: The main aspects of the LRMM model are reviewed in Sec. \ref{Sec2}, including the FCNC couplings of the neutral gauge bosons ($Z,Z^{\prime}$) and its Higgs sector. Then, we present  in Sec. \ref{Sec3} the  FCNC top quark phenomenology; here we include both the constraints  on the parameters of the model, as well as the study of the single top production through  the process $e^+e^-\to t\bar{c}+\bar{t}c$, which is mediated by the  FCNC couplings of the neutral $Z$ and $Z^{\prime}$ gauge bosons, we also discuss the signal vs background levels and determine the detectability of the signal at the CECP with $\sqrt{s}=210$ GeV. Finally, Sec. \ref{Conclusions} contains our conclusions.




\section{The Left-Right Mirror model - theoretical considerations \label{Sec2}}


Among the possible extensions of the SM are the LR models, have been widely studied, particularly because they try to explain the apparent LR asymmetry of our world. The LR model contains the electroweak group  $\rm SU(2)_L \times SU(2)_R \times U(1)_{B - L}$ and it renders the baryon-lepton number symmetry $\rm U(1)_{B-L}$ more natural by gauging it; it has also been proposed as a
solution to the strong CP problem when accompanied by the introduction of mirror fermions \cite{Barr:1991qx, Babu:1989rb}. Such an extended gauge group with additional fermions,  apart from fitting nicely into unification schemata, restores the left-right symmetry missing in the SM, in a manner that goes beyond the simplest left-right symmetry models. It is interesting to ask about
the physical implications of this model and its test at current and future colliders. So, let us present first the Lagrangian of the LRMM.


\subsection{Left-Right Mirror Model}
In the left-right model with mirror fermions (LRMM) \cite{Ceron:1997ey, Cotti:2002zq}, the
right-handed (left-handed) components of mirror fermions transform as
doublets (singlets) under the new $\rm SU(2)_R$. The SM fermions are singlets
under the $\rm SU(2)_R$, whereas the right-handed mirrors fields are also singlets
under $\rm SU(2)_L$. Mirror and SM fermions will share hypercharge and
color interactions. Thus, the fermion fields will be written as follows:

\begin{eqnarray*}
	\ell_{iL}^{o} & = & \left(\begin{array}{c}
		\nu_{i}^{o}\\
		e_{i}^{o}
	\end{array}\right)_{L},\,e_{iR}^{o},\,\nu_{iR}^{o};\;\ell_{iR}^{\prime o}=\left(\begin{array}{c}
		\nu_{i}^{\prime o}\\
		e_{i}^{\prime o}
	\end{array}\right)_{R},\,e_{iL}^{\prime o},\,\nu_{iL}^{\prime o},\\
	q_{iL}^{o} & = & \left(\begin{array}{c}
		u_{i}^{o}\\
		d_{i}^{o}
	\end{array}\right)_{L},\,u_{iR}^{o},\,d_{iR}^{o};\;q_{iR}^{\prime o}=\left(\begin{array}{c}
		u_{i}^{\prime o}\\
		d_{i}^{\prime o}
	\end{array}\right)_{R},\,u_{iL}^{\prime o},\,d_{iL}^{\prime o},
\end{eqnarray*}
where the index $i$ runs over the three fermions families and the superscript ($o$) denote weak eigenstates; the primes
will be associated with the mirror particles.The field content is summarized in table \ref{fields}.
\begin{table}[!ht]
\begin{centering}
	\begin{tabular}{cc}
		\hline
		Field $\,\,\,\,\,\,\,\,$ &  $ SU(3)_C\otimes SU(2)_L \otimes SU(2)_R \otimes U(1)_{Y^{\prime}}$  \\
		\hline
		$\ell^{o}_{iL}$ & $(\boldsymbol{1}, \;\boldsymbol{2}, \;\boldsymbol{1}, \;-1)$ \\
		$\nu^{o}_{iR}$ &  $ (\boldsymbol{1}, \;\boldsymbol{1},\;\boldsymbol{1}, \;0)$  \\
		$e^{o}_{iR} $ &  $(\boldsymbol{1}, \;\boldsymbol{1},\; \boldsymbol{1}, \;-2)$  \\
		$\nu'^{o}_{iL}$ &  $  (\boldsymbol{1}, \;\boldsymbol{1}, \;\boldsymbol{1}, \;0)$  \\
		${e'}^{o}_{iL} $ & $ (\boldsymbol{1}, \; \boldsymbol{1}, \; \boldsymbol{1},  \;-2)$ \\
		${l'}^{o}_{iR} $ &  $ (\boldsymbol{1}, \;\boldsymbol{1}, \;\boldsymbol{2},\; -1)$  \\
		\hline
		$u^{o}_{iR} $ & $(\boldsymbol{3}, \;\boldsymbol{1}, \;\boldsymbol{1}, \;4/3)$ \\
		$d^{o}_{iR}$ & $(\boldsymbol{3},  \;\boldsymbol{1},  \;\boldsymbol{1}, \; 2/3)$ \\
		${u'}^{o}_{iL}$ & $(\boldsymbol{3},  \;\boldsymbol{1},  \;\boldsymbol{1},  \;4/3)$ \\
		${d'}^{o}_{iL}$ & $ (\boldsymbol{3},  \;\boldsymbol{1},  \;\boldsymbol{1},  \;2/3)$ \\
		$q^{o}_{iL} $ & $(\boldsymbol{3}, \; \boldsymbol{2},  \;\boldsymbol{1},  \;1/3)$ \\
		${q'}^{o}_{iR}$ & $(\boldsymbol{3}, \;\boldsymbol{1},  \;\boldsymbol{2},  \;1/3)$ \\
		\hline
		$\phi$ &$(\boldsymbol{1}, \;\boldsymbol{2}, \;\boldsymbol{1}, \;-1)$ \\
		$\phi'$ &$(\boldsymbol{1}, \;\boldsymbol{1}, \;\boldsymbol{2}, \;-1)$ \\
		\hline
	\end{tabular}
	\caption{The irreducible representations are presented for fermion and scalar fields of the LRMM. The bold numbers denote the dimensions of representations of $SU(3)_C$, $SU(2)_L$ and $SU(2)_R$.}
	\label{fields}
	\par\end{centering}
\end{table}

\subsection{Scalar sector and symmetry breaking}
\label{symm-break}
The symmetry breaking is realized by including two Higgs doublets, the SM
one ($\phi$) and its mirror partner ($\phi'$). The scalar fields are parametrized as
\begin{equation}
\phi = \left( \begin{array}{c}
\phi_1^+\\  
\phi^{0}
\end{array} \right),\;
\phi^{'} = \left( \begin{array}{c}
\phi_2^+\\
\phi^{'0}
\end{array} \right).
\label{eq:scalar_reps}
\end{equation}

The vacuum expectation values (VEV's) of the Higgs field are
\begin{equation}
\langle \phi \rangle
=
\left(
\begin{array}{l}
0\\
\frac{v}{\sqrt{2}}
\end{array}
\right),
\hspace{1mm}
\langle \phi' \rangle
=
\left(
\begin{array}{l}
0 \\
\frac{v^\prime}{\sqrt{2}}
\end{array}
\right),
\label{vac}
\end{equation}
where $v=246$ GeV and the value for $v'$ must satisfies $v'> v$. The symmetry breaking pattern should be as follows $SU(2)_L \otimes SU(2)_R \otimes U(1)_{Y^{\prime}}  \xrightarrow{\langle \phi' \rangle} SU(2)_L \otimes U(1)_{Y} \xrightarrow{\langle \phi \rangle}  U(1)_{\text{EM}}.$
The most general potential that develops this pattern of VEV's is:

\begin{equation}
V
=
-\left(
\mu_1^2 \phi^\dag \phi + \mu_2^2 {\phi'}^\dag \phi'
\right)+
\frac{\lambda_1}{2}
\left(
\left(
\phi^\dag \phi
\right)^2 +
\left(
{\phi'}^\dag \phi'
\right)^2
\right)+
\lambda_2
\left(
\phi^\dag \phi
\right)
\left(
{\phi'}^\dag \phi'
\right).
\label{potentialLR}
\end{equation}
The terms with $\mu_1, \mu_2$ are included  so that the
parity symmetry (P) is broken softly, i.e. only through the
dimension-two mass terms of the Higgs potential.

After the symmetry breaking, the neutral Higgs boson squared mass matrix that follows from
this potential is:
\begin{equation}
\rm {\sf M}^2_{H^0}
=
\left(
\begin{array}{cc}
2 \lambda_1 v^2   &  2 \lambda_2 v v' \\
2 \lambda_2 v v'  &  2 \lambda_1 {v'}^2
\end{array}
\right) .
\end{equation}

Diagonalization of the Higgs-boson squared mass matrix is
straightforward using a real basis. Out of the eight scalar
degrees of freedom associated with two complex doublets, six
become the Goldstone bosons required to give mass to $W^\pm$,
${W'}^\pm$, $Z$ and $Z'$. Thus only two neutral Higgs bosons remain;
the neutral physical states $(H,\,H^{\prime})$ are related to the weak states as follows:
\begin{equation}
\begin{pmatrix}
H \\
H^\prime
\end{pmatrix}=
\begin{pmatrix}
\cos\alpha & \sin\alpha \\
-\sin\alpha & \cos\alpha
\end{pmatrix}
\begin{pmatrix}
\textrm{Re}[\phi^0] \\
\textrm{Re}[\phi^{\prime0}]
\end{pmatrix}
\end{equation}
%
%
%
%
where $\alpha$ denotes the neutral Higgs mixing angle. The scalar state $H$ is identified as the SM-like Higgs boson with $m_H=125$ GeV and $H^{\prime}$ is a new heavy scalar boson.

\subsection{ Gauge bosons masses and mixing }
The mass matrix for the gauge bosons is obtained from the kinetic terms for scalars in the Lagrangian
\begin{equation}
\rm {\cal L}_{\textrm{scalar}}
=
\left(
D^\mu \phi
\right)^\dag
\left(
D_\mu \phi
\right)
+
\left(
{D'}^\mu \phi'
\right)^\dag
\left(
D'_\mu \phi'
\right)
\label{boson}
\end{equation}
where $D_\mu$ denotes the covariant derivative associated with the
SM, and $D'_\mu$ is the one associated with the mirror sector, which are written as
\begin{eqnarray}
D_\mu=\partial_\mu+ig_2\frac{\vec{\tau}}{2}\cdot \vec{W}_{L\mu}+ig_1Y B_\mu,\\ 
D^\prime_\mu=\partial_\mu+ig_2^\prime\frac{\vec{\tau}}{2}\cdot \vec{W}_{R\mu}+ig_1Y^\prime B_\mu,
\end{eqnarray}
where $g_2$, $g'_2$ and $ g_1$ are the coupling constants
associated with the $\rm SU(2)_L$, $\rm SU(2)_R$ and $\rm U(1)_{Y^\prime}$ gauge groups,
respectively. The hypercharge $Y$ is related with $Y^\prime$ as $Y/2= T_{3R}+Y^\prime/2$.
After substituting the VEV's from Eq.(\ref{vac}) in the Lagrangian given  by Eq. (\ref{boson}), we obtain the expressions for the mass matrices of the gauge bosons. The mass matrix for the charged gauge bosons is already diagonal, with mass eigenvalues: $ M_W = \frac{1}{2} vg_2$ and $M_{W'}= \frac{1}{2} v'g'_2$. Meanwhile, the mass matrix for the neutral gauge bosons is not diagonal,
and it is given by
\begin{equation}
M_0^2= \frac{1}{4}
\left(
\begin{array}{ccc}
g^2_2 v^2        & 0             &  -g_2  g_1 v^2 \\
0            & {g'_2}^2 {v'}^2 &  -g'_2 g_1 {v'}^2\\
-g_2 g_1 v^2  & g'_2 g_1 {v'}^2 &  g_1^2 \left( v^2 + {v'}^2
\right)
\end{array}
\right) .
\end{equation}

However this matrix can be diagonalized by an orthogonal transformation
$\sf R$, which relates the weak and mass eigenstates, and is given by
{
	\begin{equation}
	\sf R = \left(
	\begin{array}{ccc}
	c_{\theta_w}  c_{\Theta} & c_{\theta_w}  s_{\Theta} &
	s_{\theta_w}\\
	-\frac{1}{c_{\theta_w}} \left( s_{\Theta} r_{\theta_w} +
	\frac{g_2}{g'_2} c_{\Theta} s^2_{\theta_w} \right) &
	\frac{1}{c_{\theta_w}} \left( c_{\Theta} r_{\theta_w} -
	\frac{g_2}{g'_2} s_{\Theta} s^2_{\theta_w} \right) &
	\frac{ g_2}
	{g'_2} s_{\theta_w} \\
	t_{\theta_w} \left( \frac{ g_2}{g'_2} s_{\Theta} -
	r_{\theta_w} c_{\Theta} \right) & - t_{\theta_w} \left( \frac{
		g_2}{g'_2} c_{\Theta} + r_{\theta_w} s_{\Theta} \right) &
	r_{\theta_w}
	\end{array}
	\right),
	\label{rot}
	\end{equation}}
with $\theta_w$ and $\Theta$ denoting the rotation angles of the
neutral gauge bosons and $r_{\theta_w} \equiv \sqrt{c^2_{\theta_w} -
	\frac{g^2_2}{{g'}^2_2} s^2_{\theta_w}}$.
Once the rotation is done, it is found one massless state (the photon) and two massive states, with
eigenvalues given by
%
%
{
	\begin{eqnarray}
		M_{Z,Z'} & = & \frac{1}{8}\left[v^{2}\left(g_{2}^{2}+g_{1}^{2}\right)+{v'}^{2}\left({g'}_{2}^{2}+{g'}_{1}^{2}\right)\right]\nonumber\\
		& \mp & \frac{1}{8}\sqrt{\left[v^{2}\left(g_{2}^{2}+g_{1}^{2}\right)+{v'}^{2}\left({g'}^{2}+g_{1}^{2}\right)\right]^{2}-4v^{2}{v'}^{2}\left(g_{2}^{2}g_{1}^{2}+g_{2}^{2}{g'}_{2}^{2}+{g'}_{2}^{2}g_{1}^{2}\right)}.
	\end{eqnarray}
}

%

In order to find the couplings of the Higgs fields
with the neutral gauge boson,
we expand the expressions for $D_{\mu}$ and
$D'_\mu$, substituting the physical states
into Eq. (\ref{boson}). We obtain the following expression for
the $ZZ\Phi$ ($\Phi=H,\,H^{\prime}$) interactions:
\begin{eqnarray}
	\mathcal{L}_{ZZ\Phi} & = & \sqrt{2}[g_{2}M_{W}X(\Theta,\theta_{w})\cos\alpha
	 +  g_{2}^{\prime}M_{W^{\prime}}Y(\Theta,\theta_{w})\sin\alpha]HZ_{\mu}Z^{\mu}\nonumber\\
	& + & \sqrt{2}[-g_{2}M_{W}X(\Theta,\theta_{w})\sin\alpha
	+ g_{2}^{\prime}M_{W^{\prime}}Y(\Theta,\theta_{w})\cos\alpha]H^{\prime}Z_{\mu}Z^{\mu},
\end{eqnarray}
where
\begin{equation}
X
\left(
\Theta, \theta_w
\right)
=
\left[
c_{\theta_w} c_\Theta - \frac{g_1}{g_2} t_{\theta_w}
\left(
\frac{g_2}{g'_2} s_\Theta - r_{\theta_w} c_\Theta
\right)
\right]^2,
\end{equation}

\begin{eqnarray}
	Y\left(\Theta,\theta_{w}\right)  =  \left[-\frac{1}{c_{\theta_{w}}}\left(s_{\Theta}r_{\theta_{w}}+\frac{g_{2}}{g'_{2}}c_{\Theta}s_{\theta_{w}}^{2}\right)\right.
	-  \left.\frac{g_{1}}{g_{2}}t_{\theta_{w}}\left(\frac{g_{2}}{g'_{2}}s_{\Theta}-r_{\theta_{w}}c_{\Theta}\right)\right]^{2}.
\end{eqnarray}

As far as the coupling between neutral scalars and $W^\pm$ boson is concerned, it is given by:
%
\begin{eqnarray}
{\cal L}_{W^{\pm} HH^\prime}=
\sqrt{2} g_2 M_W g^{\mu \nu}
(
H \cos\alpha - H^{\prime} \sin\alpha
)
W^-_\mu W^+_\nu.
\end{eqnarray}

\subsection{Yukawa Lagrangian and fermion mixing}
\label{yuklagr}

The renormalizable and gauge invariant interactions of the scalar
doublets $\phi$ and $\phi'$ with the leptons are described by the
Yukawa Lagrangian, which takes the form
\begin{equation}
{\cal L}^\ell_Y = 
\lambda_{i j} \overline{\ell}^o_{i L} \phi e^o_{j R} +
\lambda'_{i j} \overline{\ell'}^o_{i R} \phi' {e'}^o_{j L} +
\mu_{i j} \overline{e'}^o_{i L} e^o_{j R} + h. c.
\end{equation}
where $ i, j = 1, 2, 3$ are flavor indices that denote a summation when they are repeated. The $\lambda_{ i j}$ and
$\lambda_{ i j}^\prime$ are the matrices usually known as Yukawa matrices. Note that the gauge invariance allows the mixing terms between SM and mirror singlets. The matrix $\mu_{ i j}$ as well as matrices $\lambda_{ i j}$ and
$\lambda_{ i j}^\prime$ are arbitrary up to this point. Analogously, the corresponding Yukawa Lagrangian for the quarks fields are written as
%

\begin{equation}
{\cal L}^q_Y
=
\lambda^d_{ij}  \overline{Q}^o_{iL} \phi d^o_{jR} +
\lambda^u_{ij}  \overline{Q}^o_{iL} \tilde{\phi} u^o_{jR} +
{\lambda'}^d_{ij} \overline{Q'}^o_{iR} \phi' {d'}^o_{jL} 
+ {\lambda'}^u_{ij} \overline{Q'}^o_{iR} \tilde{\phi'} {u'}^o_{jL} +
\mu^d_{ij}      \overline{d'}^o_{iL} d^o_{jR} +
\mu^u_{ij}      \overline{u'}^o_{iL} u^o_{jR} + h.c.,
\end{equation}
where the conjugate fields $\tilde{\phi}$ ($\tilde{\phi'}$) are
obtained as $\tilde{\phi} =  i \tau_2 \phi^*$. The VEV's of the neutral scalars produce the fermion mass terms, which
in the gauge eigenstate basis read
\begin{equation}
{\cal L}_{\rm mass}
=
\overline{\psi^o_L} {\sf M} \psi^o_R + h.c. .
\end{equation}
For the lepton sector, the non-diagonal mass matrix $\sf M$,
takes the form
\begin{equation}
\sf M
=
\left(
\begin{array}{cc}
\sf K   & \sf 0       \\
\sf \mu & \sf K'
\end{array}
\right),
\end{equation}
where $\sf K = \frac{1}{2} {\lambda} v$ and $\sf K' =
\frac{1}{2} \lambda' v'$ correspond to the $3 \times 3$
matrices generated from the symmetry breaking VEV's; $\mu$
corresponds to the gauge invariant $3 \times 3$ mixing terms between
ordinary and mirror fermions singlets.
Thus, the mass matrices can be diagonalized through unitary matrices ${\sf U}_{a}$, with $a=L,R$, as follows
%
%
\begin{equation}
{\sf M}_D = {\sf U}^\dag_L {\sf M} {\sf U}_R .
\label{fermass}
\end{equation}
The mass and unitary matrices contain the information for the SM and mirror fermions. We choose to write ${\sf U}_{a}$ as
\begin{equation}
{\sf U}_a=
\begin{pmatrix}
{\sf A}_a &{\sf E}_a\\
{\sf F}_a &{\sf G}_a
\end{pmatrix}.
\label{unit}
\end{equation}
Note that in general the submatrices ${\sf A}_a$, ${\sf E}_a$, ${\sf F}_a$ and ${\sf G}_a$ are not unitary.

With the help of the relations (7-10), and working with the
Higgs mass-eigenstate basis, we can obtain the tree-level interactions of the
neutral Higgs bosons $H$ and $H^{\prime}$ with the light (SM) fermions, which are given by
\begin{eqnarray}\label{YukawaInteractions}
{\cal L}_Y^l
=
\frac{g_2}{2 \sqrt{2}}
\overline{f}_L^i ({\sf A}^\dag_L  {\sf A}_L)_{ij} \frac{m_l}{M_W} f_{R}^j
\left(
H \cos\alpha - H' \sin\alpha
\right)
+
\frac{g'_2}{\sqrt{2}}
\overline{f}_L^i \frac{m_l}{M_{W'}}
({\sf F}^\dag_R {\sf F}_R)_{ij} f_R^j
\left(
H \sin\alpha + H' \cos\alpha
\right) + h.c.
\end{eqnarray}
with $f$ denoting the up-type, down-type quarks and charged leptons. 
One can see that the couplings are not diagonal in general, thus new
FCNC interactions will be present in this model. The
resulting phenomenological constraints and predictions will be discussed
in the next sections.

Finally, once we have obtained the quark and lepton mass eigenstates,
their gauge interaction can be obtained from the
Lagrangian
\begin{equation}
{\cal L}^{\rm int}
=
\overline{\psi}  i \gamma^\mu D^\mu \psi +
\overline{\psi^{\prime}} i \gamma^\mu D'_\mu \psi^{\prime},
\label{inter}
\end{equation}
where $\psi$, $\psi'$ denote the standard and mirror fermions,
respectively.

The neutral current term for the multiplet $\psi$ of a given
electric charge, including the contribution of the neutral gauge
boson mixing, can be written as follows
\begin{equation}
-{\cal L}^{\rm nc}
=
\sum_{a =  L, R} \overline{\psi}_a^o \gamma^\mu
\left(
g_2 {\sf T}_{3 a}, g'_2 {\sf T}'_{3 a}, g_1 \frac{{\sf Y}_a}{2}
\right)
\psi_a^o
\left(
\begin{array}{l}
W^3\\
{W'}^3\\
B
\end{array}
\right)_\mu.
\end{equation}

Using Eqs.~(\ref{fermass},\ref{unit}),
one arrives to the following expression
in terms of the mass eigenstates:
\begin{equation}\label{ZZpAFermions}
\small{-{\cal L}^{\rm nc}
	=
	\sum_{a = L,R} \overline{\psi}_a \gamma^\mu {\sf U}_a^\dag
	\left(
	g_2 {\sf T}_{3a}, g'_2 {\sf T'}_{3a}, g_1 \frac{{\sf Y}_a}{2}
	\right)
	{\sf U}_a \psi_a
	\left(
	\begin{array}{l}
	Z\\
	Z'\\
	A
	\end{array}
	\right)_\mu}.
\end{equation}
where ${\sf T}_{3a}$, ${\sf T'}_{3a}$, Y are the generators of the
$\rm SU(2)_L$, $\rm SU(2)_R$, and U(1), respectively. The vertices $Ztc$ and $Z^{\prime}tc$ can be obtained from \ref{ZZpAFermions}, resulting in the following expressions: 
\begin{eqnarray}
Ztc&=&-i\frac{e}{s_Wc_W}(g_{V_{tc}}+g_{A_{tc}}\gamma_5)\gamma_{\mu}\label{Ztc},\quad Z^{\prime}tc=Ztc\left(g_{\left(V,\,A\right)_{tc}} \rightarrow g_{\left(V,\,A\right)_{tc}}^{\prime}\right), \\
g_{A_{tc}, V_{tc} } & = & \frac{1}{4}\left(\cos\Theta-\frac{\sin\Theta s_{W}^{2}}{r_{W}}\right)\left(\eta_{L}\right)_{tc}\pm\frac{1}{4}\text{\ensuremath{\frac{c_{W}^{2}\sin\Theta}{r_{W}}}}\left(\lambda_{R}\right)_{tc},\\
g^{\prime}_{A_{tc}, V_{tc} } & = & \frac{1}{4}\left(\sin\Theta+\frac{\cos\Theta s_{W}^{2}}{r_{W}}\right)\left(\eta_{L}\right)_{tc}\mp\frac{1}{4}\text{\ensuremath{\frac{c_{W}^{2}\cos\Theta}{r_{W}}}}\left(\lambda_{R}\right)_{tc},
\end{eqnarray}
where $s_W$ and $c_W$ are the sine and cosine of the weak angle. While the couplings $Htc$ and $H^{\prime}tc$ are given in Eq. \eqref{YukawaInteractions}.

There are several physical consequences from these interactions. In particular, one can identify the presence of
Lepton Flavor Violation, which provides  an interesting signal to be searched at LHC and future colliders, 
in particular the decay mode $h\to\tau\mu$, which has been studied in the literature
\cite{DiazCruz:1999xe, Arroyo-Urena:2018mvl, Arroyo-Urena:2020mgg, Hernandez-Tome:2020lmh, Han:2000jz, Vicente:2019ykr, Lami:2016mjf, DiazCruz:2008ry}.  Within the LRMM the possibility to search for LFV Higgs interactions at LHC
 was studied in ref. \cite{Cotti:2002zq}.


\section{Phenomenological analysis for $e^+e^- \to tc$\label{Sec3}}


\subsection{Constraint on the free LRMM parameters}
In order to perform a realistic numerical analysis of signal and main backgrounds, we need to analyze the free parameter of the model that has an impact on the FCNC signal of our interest. Then one can define the allowed parameter space  involved in subsequent calculations.  Among those parameters we shall consider
the following set:
\begin{enumerate}
	\item Mixing angle between the neutral gauge boson $Z$ and $Z^{\prime}$: $\Theta$,
	\item Mixing angle of the Higgs sector: $\alpha$,
	\item The matrix elements that will induce FCNC interactions:
	\begin{enumerate}
		\item $\left(A^{\dagger}_{L} A_{L}\right)_{f_if_j}\equiv \left(\eta_{L}\right)_{f_if_j}$,
		\item $\left(F^{\dagger}_R F_R\right)_{tc}\equiv \left(\lambda_R\right)_{tc}$,
	\end{enumerate} 
	\item The $Z^{\prime}$ gauge boson mass, $m_{Z^{\prime}}$.
\end{enumerate}

\begin{figure}[!ht]
	\centering
	\includegraphics[scale=0.5]{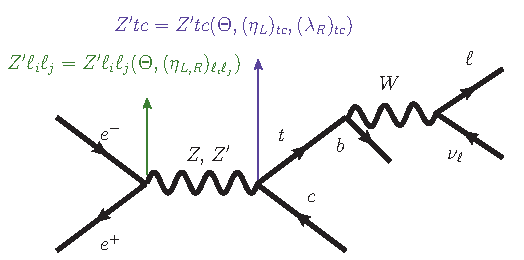}
	\caption{Feynman diagrams that contribute to the process $e^-e^+\to V \to tc\;(\to \ell \nu_{\ell} b c)$ with $V=\,Z,\,Z^{\prime}$ and $\ell=e,\,\mu$.
	}
	\label{FD_eetc_LRSM}
\end{figure}

The matrix element $\left(\eta_{L}\right)_{tc}$, along with $\left(\lambda_R\right)_{tc}$ and $\Theta$, 
determine the flavor-changing couplings $Z(Z^{\prime})tc$ and they are necessary for our study, as can be seen in Fig. \ref{FD_eetc_LRSM} whose Feynman diagrams represent the signal we are searching for. There are also contributions of the SM-like Higgs boson and its heavy partner $H^{\prime}$, but those are negligible.  

We first analyze the constraints on the mixing angle $\alpha$ by using the Higgs boson signal strengths. For a production process $\sigma(pp\to \phi)$ and decay $\phi\to X$, the corresponding signal strength is defined as follows:

\begin{equation}
\mathcal{R}_{X}=\frac{\sigma(pp\to h)\cdot\mathcal{BR}(h\to X)}{\sigma(pp\to h^{SM})\cdot\mathcal{BR}(h^{SM}\to X)},
\end{equation}
where $\sigma(pp\to \phi)$ being the production cross section of $\phi=h,\,h^{SM}$; here $h$ is identified as the SM-like Higgs boson coming from an extension of the SM and $h^{\text{SM}}$ is the SM Higgs boson; $\mathcal{BR}(\phi\to X)$ is the branching ratio of the decay $\phi\to X$, with $X=b\bar{b},\;\tau^-\tau^+,\;\mu^-\mu^+,\;WW^*,\;ZZ^*,\;\gamma\gamma$.

\begin{figure}[!ht]
	\centering
	\subfigure[]{\includegraphics[width=0.37\textwidth]{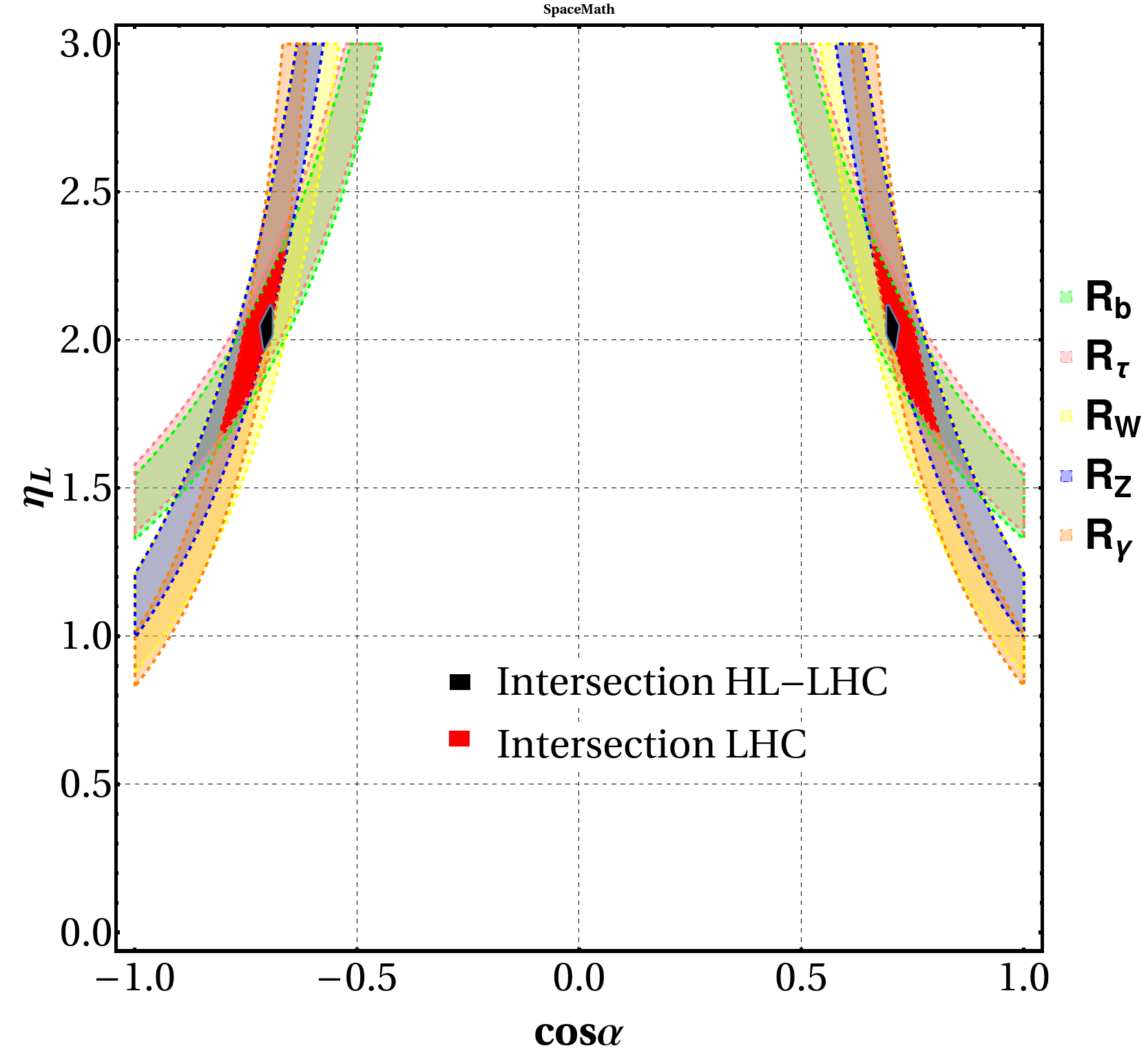}\label{calphaVSetaL}}
	\subfigure[]{\includegraphics[width=0.32\textwidth]{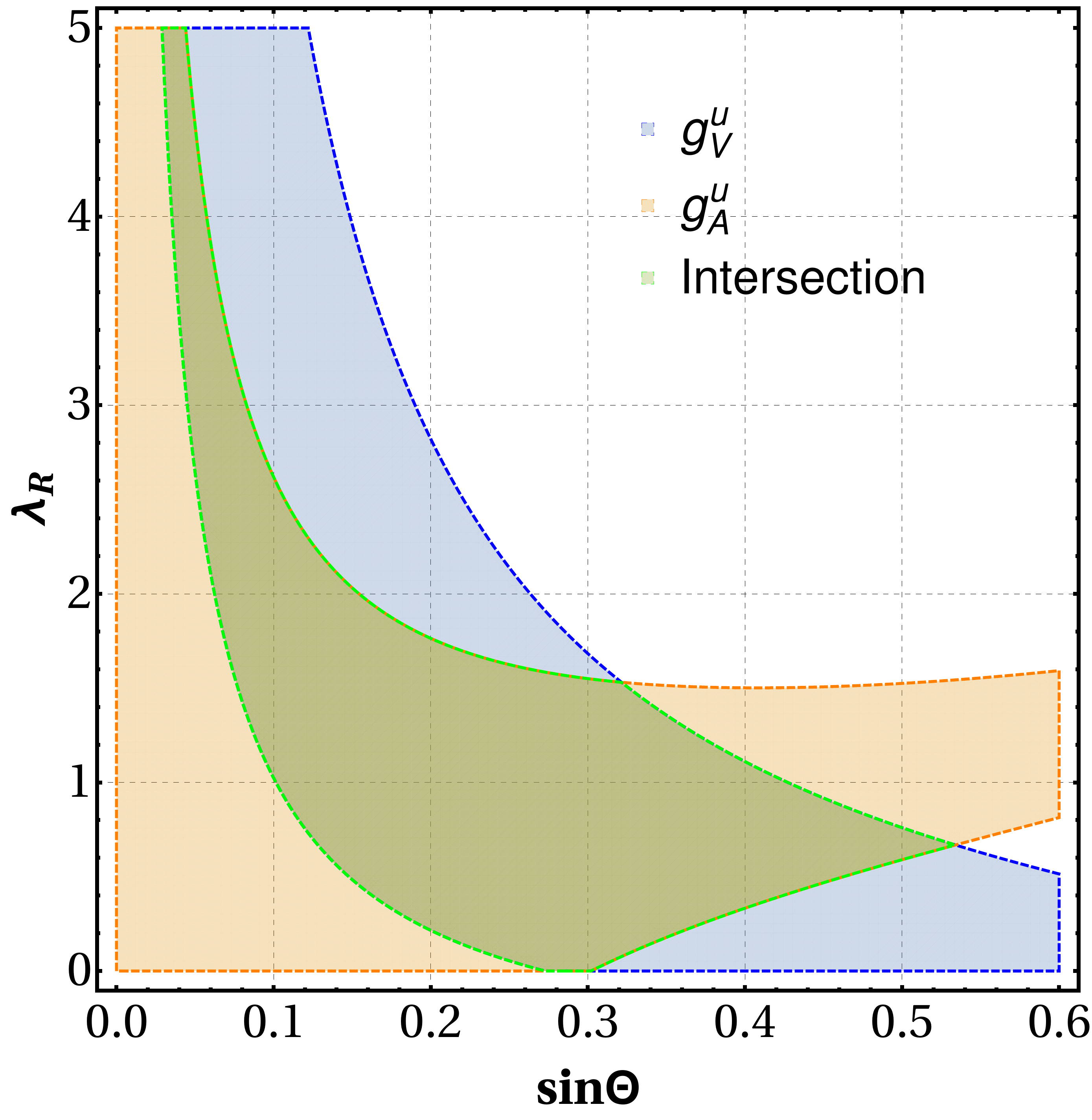}\label{lamdRVSsinTheta}}
	\caption{(a) $\cos\alpha-\eta_L$ plane. The green (pink, yellow, blue, orange) area represents the allowed region by $\mathcal{R}_b$ ($\mathcal{R}_{\tau}$, $\mathcal{R}_W$, $\mathcal{R}_Z$, $\mathcal{R}_{\gamma}$), while the red area is the intersection of all individual $\mathcal{R}_{X}'s$ considering experimental results of the LHC; finally, the black area corresponds to the expected measurements at the HL-LHC. (b) $\sin\Theta-\lambda_R$ plane. The orange and blue areas are the allowed regions by the axial and vector-axial parts of the $Zq_i^u\bar{q}_i^u$ coupling, respectively. Meanwhile, the intersection of them is the green region.} 	
\end{figure}

Figure \ref{calphaVSetaL} shows the $\cos\alpha$-$(\eta_L)_{ii}\footnote{Here $i=b, t,\tau$. Assuming it does not drastically change our results since the dominant term is that of the SM.}$ plane with the allowed region in red color obtained by the intersection of $\mathcal{R}_{b,\tau,W,Z,\gamma}$, whose values were taken from the reports of the CMS and ATLAS collaborations\cite{CMS:2020gsy, ATLAS:2019nkf}, while the black area corresponds to the expected measurements at the High Luminosity LHC (HL-LHC)\cite{Cepeda:2019klc}; from here on we will omit the subscripts $(ii)$ so as not to overload the notation unless otherwise stated. We note that $\eta_L$ can take values from about 1.7 to 2.3 (LHC) and 1.97 to 2.1 (HL-LHC) depending on $\cos\alpha$. For the particular case when $\cos\alpha=\pm 0.7$, we obtain $2\lesssim\eta_L\lesssim 2.2$ (LHC) and $2\lesssim\eta_L\lesssim 2.1$ (HL-LHC). The plots were generated with the \texttt{SpaceMath} package \cite{Arroyo-Urena:2020qup}.
           
           Once the allowed values for $\eta_L$ are obtained, we take a particular value $\eta_L=2.1$ for a fixed value of $\cos\alpha=0.7$ (in accordance with both LHC and HL-LHC), then we analyze the $Z$ gauge boson couplings to quarks of type up: $g_V^u$ and $g_A^u$, in order to have an allowed region for the parameters $(\lambda_R)_{jj}$ and $\Theta$. As in the case of the matrix elements of $\eta$, for simplicity we will omit the subscripts that represent the matrix elements of $\lambda_R$. Figure \ref{lamdRVSsinTheta} shows the $\sin\Theta-\lambda_R$ plane in which the green area represents the intersection of the consistent regions with the reported values for $g_V^u$ (blue area) and $g_A^u$ (orange area) \cite{Zyla:2020zbs}. We notice that for $\sin\Theta\approx 0.05$, $2.5 \lesssim\lambda_R\lesssim 4.7$; and for $\sin\Theta=0.5$, $0.62 \lesssim\lambda_R\lesssim 0.75$; the general behavior is that
$\lambda_R$ grows as $\sin\Theta$ increases and vice versa.

\begin{figure}[!ht]
	\centering
	\includegraphics[width=0.35\textwidth]{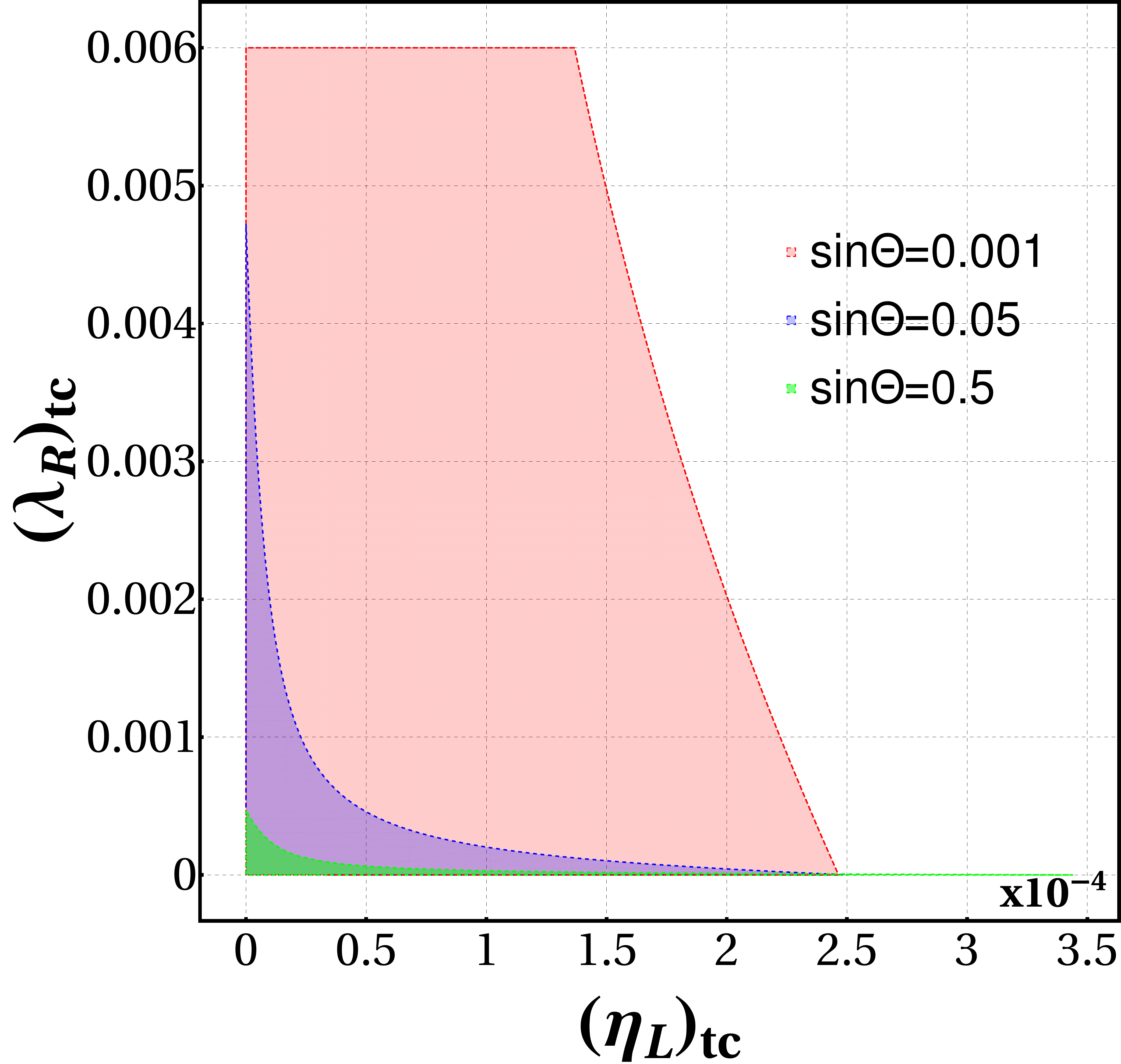}
	\caption{The colored areas in the $\left(\eta_L\right)_{tc}-\left(\lambda_R\right)_{tc}$ plane represent the allowed regions by the upper limit on the $\mathcal{BR}(t\to Zc)$ expected at HL-LHC. Green area: $\sin\Theta=0.5$, purple area: $\sin\Theta=0.05$, red area: $\sin\Theta=0.001$ 
	}
	\label{EtaL-LambR} 
\end{figure}

As far as the matrix elements $\left(\eta_{L}\right)_{tc}$ and $\left(\lambda_R\right)_{tc}$ are concerned, these can be constrained through the upper limit of the $\mathcal{BR}(t\to Z c)$. It is expected for CEPC to have a sensitivity for the $\mathcal{BR}(t\to Z c)$ decay of order $10^{-4}$, instead of that value, we take the expected upper limit on $\mathcal{BR}(t\to Z c)\lesssim5\times 10^{-6}$ at HL-LHC \cite{ATLAS:2016qxw}.
Thus, we present in Fig. \ref{EtaL-LambR} the $(\eta_L)_{tc}-(\lambda_R)_{tc}$ plane which shows the allowed region by the expected upper bound on $\mathcal{BR}(t\to Z c)$.     
We observe, in a general way, that the behavior of $\left(\lambda_R\right)_{tc}$ is increasing as $\left(\eta_L\right)_{tc}$ decreasing.

On the other hand, to constrain the  $Z^{\prime}$ gauge boson mass, we now turn to analyze the $Z^{\prime}$ production cross-section times the branching ratio of $Z^{\prime}$ decaying into $\ell^-\ell^+$, with $\ell=e,\,\mu$. The ATLAS collaboration \cite{Aad:2019fac} searched for a new resonant high-mass phenomena in dilepton final states at $\sqrt{s}=13$ TeV with an integrated luminosity of 139 fb$^{-1}$. Nevertheless no significant deviation from the SM prediction was observed. However, lower limit excluded on the resonant mass was reported, depending on specific models.

\begin{figure}[!ht]
	\centering
	\includegraphics[angle=270,width=0.35\textwidth]{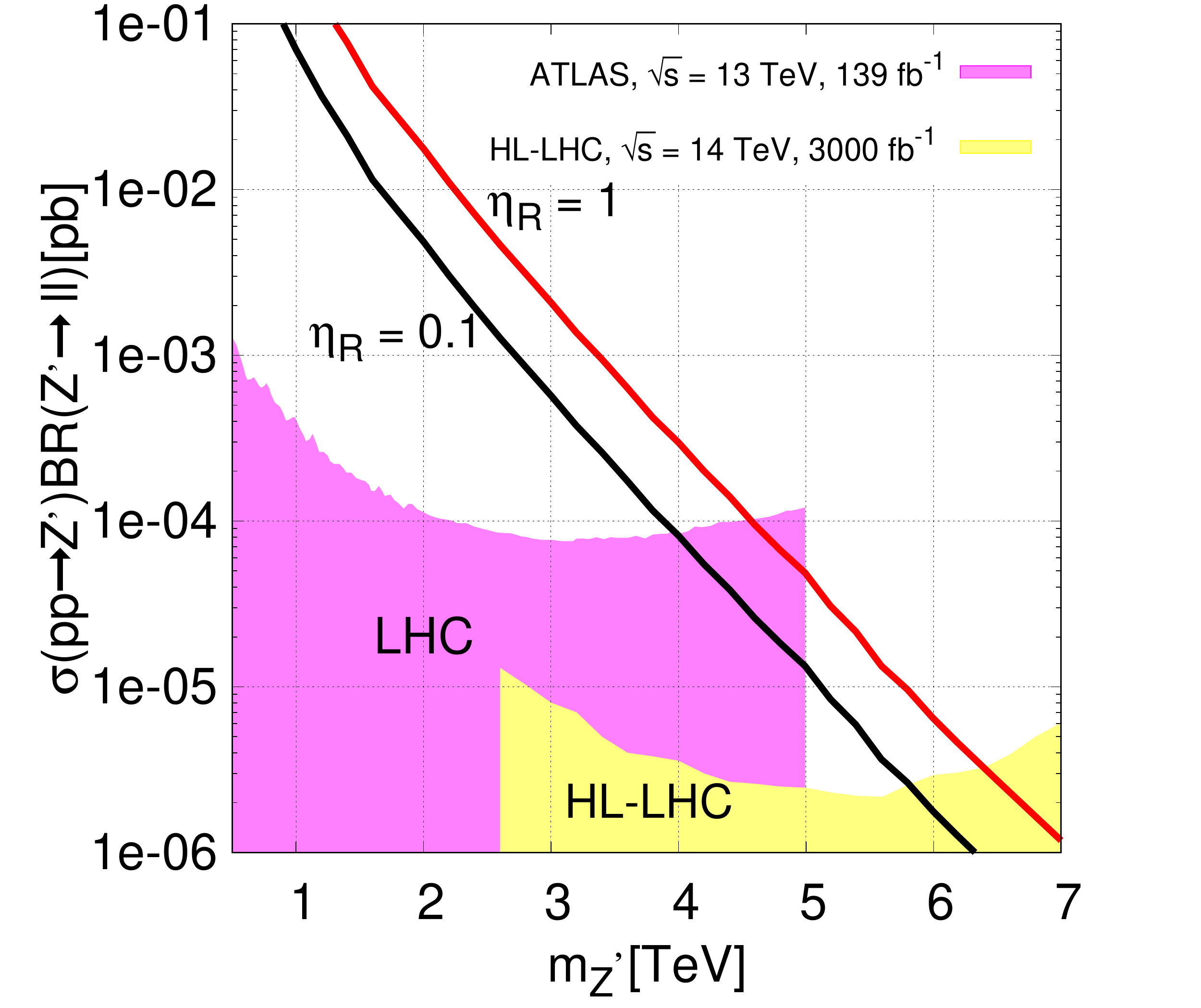}
	\caption{
		$\sigma(pp\to Z^{\prime})\mathcal{BR}(Z^{\prime}\to\ell\ell)$ as a function of $Z^{\prime}$ gauge boson mass for $\eta_{R}$ = 0.1 (black line) and $\eta_{R}$ = 1 (red line).
	}
	\label{XSZp_mZp} 
\end{figure}

Fig. \ref{XSZp_mZp} shows $\sigma(pp\to Z^{\prime})\mathcal{BR}(Z^{\prime}\to\ell\ell)$ (with $\ell=e,\,\mu$) as a function of the $Z^{\prime}$ gauge boson mass for $\eta_{R}$ = 0.1 and 1. We only include the expected limit with 95\% of Confidence Level which is represented by the upper border of the magenta area reported by ATLAS collaboration \cite{Aad:2019fac} on $\sigma(pp\to Z^{\prime})\mathcal{BR}(Z^{\prime}\to\ell\ell)$ in the combined electron and muon channels. While the yellow area represents the same but for expected measurements at the HL-LHC \cite{ATLAS:2018tvr}.
We find for the LHC (HL-LHC) $m_{Z^{'}}\lesssim 4$ ($m_{Z^{'}}\lesssim 5.8$)  TeV for $\eta_{R}=0.1$ and $m_{Z^{'}}\lesssim 4.6$ ($m_{Z^{'}}\lesssim 6.4$) TeV for $\eta_{R}=1$ are excluded. This result is in agreement with the limits reported in Ref. \cite{Zyla:2020zbs}. 

We summarize in Table \ref{ParamValues} the setting values of the LRMM parameters used in our calculation. Besides, for further analysis, we define the following scenarios:
\begin{itemize}
  \item (a) $\sin\Theta=0.001$, $(\lambda_R)_{tc}=0.003,\,(\eta_L)_{tc}=0.00018$,
	\item (b) $\sin\Theta=0.05$, $(\lambda_R)_{tc}=0.00065,\,(\eta_L)_{tc}=0.00003$,
	\item (c) $\sin\Theta=0.5$, $(\lambda_R)_{tc}=0.00002,\,(\eta_L)_{tc}=0.00014$.
\end{itemize} 
	
\begin{table}
	
	\caption{Values of the LRMM parameters used in our calculations. The subscript $f$ is understood by a lepton ($\ell=e,\,\mu$) or by a quark ($q=u,\,d,\,c,\,s,\,b$).  \label{ParamValues}}
	
	\begin{centering}
		\begin{tabular}{cc}
			\hline 
			Parameter & Value\tabularnewline
			\hline 
			\hline 
			$\left(\eta_L\right)_{tc}$ & 0.00018, 0.00003, 0.00014 \tabularnewline
			\hline 
			$\left(\lambda_R\right)_{tc}$ & 0.003, 0.00065, 0.00002 \tabularnewline
			\hline 
			$\sin\Theta$ & 0.001, 0.05, 0.5 \tabularnewline
			\hline 
			$m_{Z^{\prime}}$ & 7 TeV\tabularnewline
			\hline
			$(\eta_{R})_{ff}$ & 0.1, 1\tabularnewline
			\hline 
		\end{tabular}
		\par\end{centering}
\end{table}

\subsection{Single top FCNC production through  $e^-e^+\to tc$ for LRMM}
The process $e^-e^+\to tc$ receives contributions at tree level from gauge $Z$ and $Z^{\prime}$ bosons, $H$ and $H^{\prime}$. The Feynman diagram is shown in Fig. \ref{FD_eetc_LRSM}. The contributions of the SM-like Higgs boson $H$ and the heavy scalar $H^{\prime}$, are highly suppressed; so from now on, we will not consider them. 

\begin{figure}[!ht]
	\centering
	\includegraphics[scale=0.3]{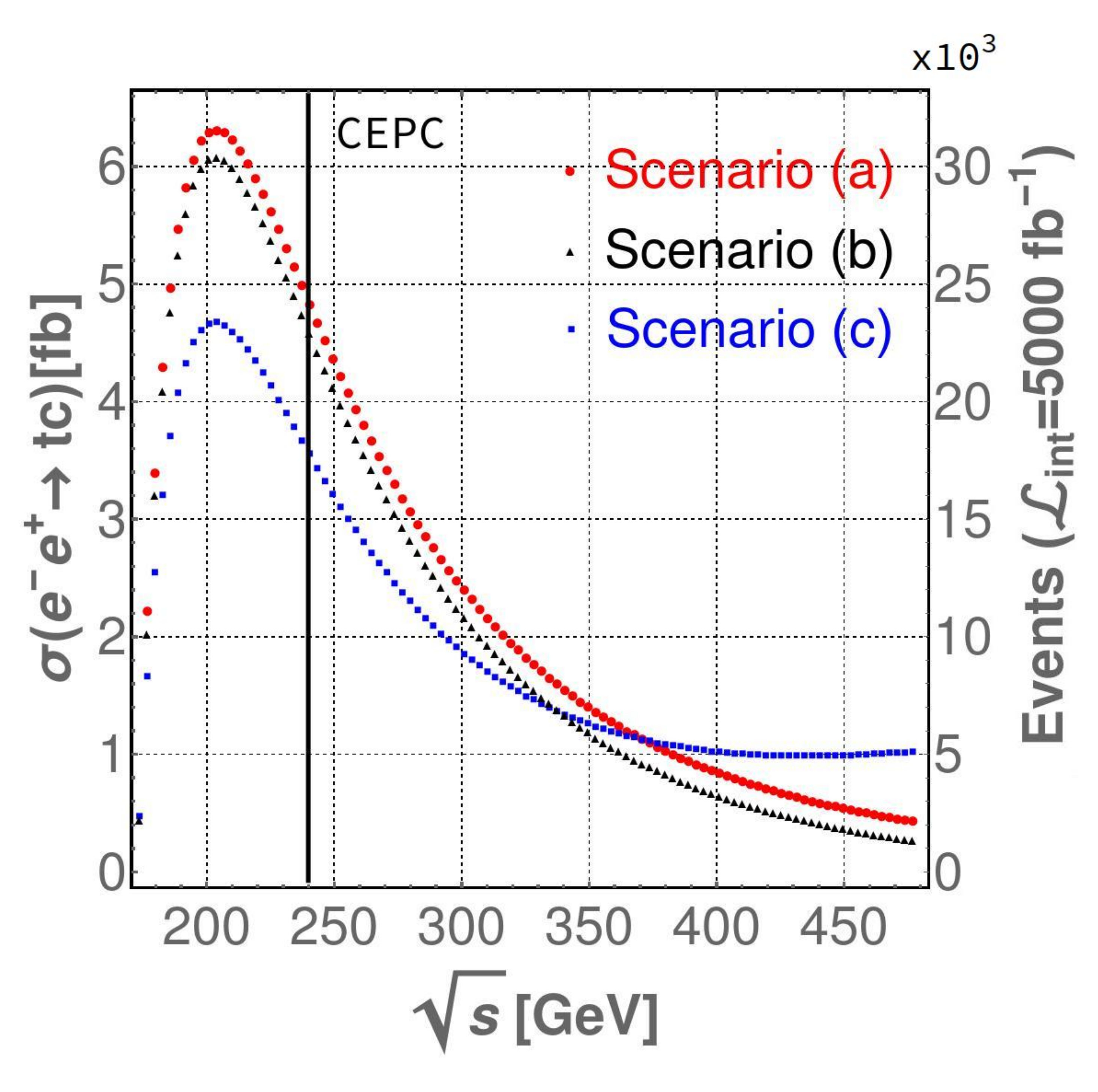}
	\caption{Production cross section of the top-charm quark pair at CEPC as a function of the center-of-mass energy for scenarios (a), (b) and (c).
	}
	\label{XS-energy}
\end{figure}
\begin{figure}[!ht]
	\centering
	\includegraphics[width=0.45\textwidth]{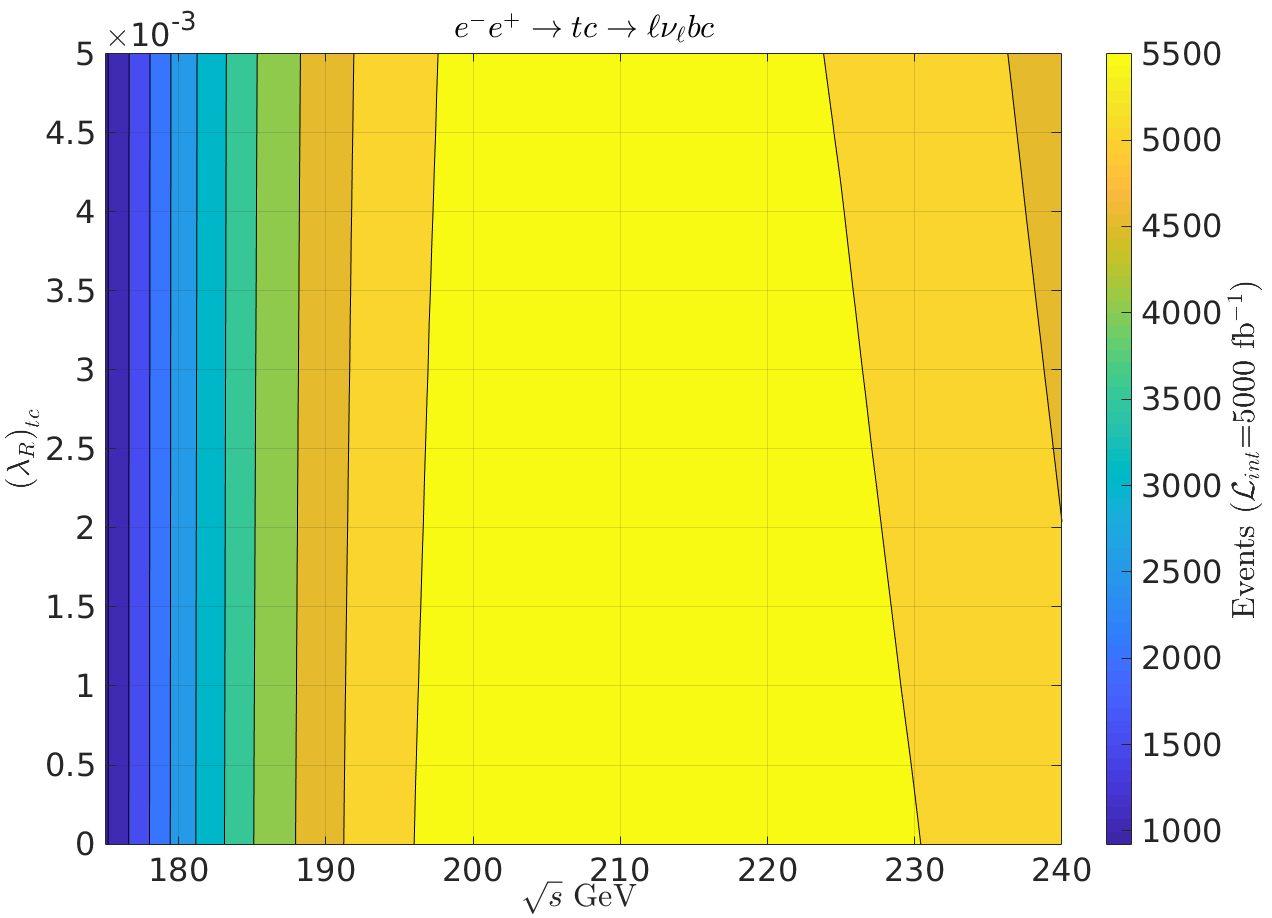}
	\caption{Produced number of signal events as a function of $\sqrt{s}$ and $\left(\lambda_R\right)_{tc}$
		with $\mathcal{L}_{\text{int}}=5000$ fb$^{-1}$ and $\left(\eta_{L}\right)_{tc}=0.00005$.	
	}
	\label{EventsLRSM} 
\end{figure}

The dominant contribution of the single top FCNC production is through a virtual $Z$ gauge boson, while the contribution of the $Z^{\prime}$ boson is up to 6 orders of magnitude lower. Although we have included both contributions in the simulation, we only present the contribution of the $Z$ gauge boson in the production of the $tc$ quark pair:

	\begin{eqnarray}
	 \sigma(e^{-}e^{+}\to Z^{*}\to tc)&=&\frac{3\pi\alpha}{c_{W}^{4}s_{W}^{4}\left(m_{Z}^{2}-s\right)^{2}}\left[(8s_{W}^{4}-4s_{W}^{2}+1)\right.
	 \nonumber\\&+&  g_{A_{tc}}^{2}(m_{c}^{2}(2m_{t}^{2}-s-2t)-2m_{c}m_{t}s-m_{t}^{2}(s+2t)+s^{2}+2st+2t^{2})\nonumber\\&+& \left.g_{V_{tc}}^{2}(m_{c}^{2}(s+2t-2m_{t}^{2})-2m_{c}m_{t}s+m_{t}^{2}(s+2t)-s^{2}-2st-2t^{2})\right]
	\end{eqnarray}

with
\begin{eqnarray*}
		t&=&(p_{e^-}-p_{top})^{2}=(p_{e^+}-p_{c})^{2}, \; s=(p_{e^-}+p_{e^+})^{2}=(p_c+p_{top})^{2}. 
\end{eqnarray*}
We present on the left axis of Fig. \ref{XS-energy} a general view of the cross section $\sigma(e^-e^+\to tc)$ as a function of the center-of-mass energy from $\sqrt{s}=m_t+m_c$ to 500 GeV for scenarios (a), (b) and (c). Meanwhile, on the right axis we show the number of events produced considering an integrated luminosity of 5000 fb$^{-1}$. For a center-of-mass energy about 210 GeV, it is observed the maximum number of events  for all scenarios. 
We note that the proposed energy regime for CEPC encompasses the largest cross sections, with scenario (a) dominating over the other ones. Besides, scenario (c) it is dominant in a regime of energy larger than $\sqrt{s}\sim 370$ GeV which motivates their analysis for $e^+e^-$ colliders with higher energy than CEPC. Nevertheless, scenario (a) is the more favored by $Z-Z^{\prime}$ mixing \cite{Osland:2020onj}, while scenario (b) agrees with \cite{Gaitan:2004by}.
Fig. \ref{EventsLRSM} shows a general overview of the behavior of signal events but for the final state, i.e, $\ell\nu_{\ell}bc$, with $\ell=e,\,\mu$; we show it as a function of $\left(\lambda_{R}\right)_{tc}$ and $\sqrt{s}$ for a representative integrated luminosity of 5000 fb$^{-1}$ and $\left(\eta_{L}\right)_{tc}=0.00005$.

We note that for specific zones of the $\sqrt{s}$-$\left(\lambda_R\right)_{tc}$ plane, up to $5500$ events could be searched. To achieve an isolated signal, we performed a Monte Carlo simulation in which we included the main SM background, as described below.

\subsubsection{Signal vs backgrounds study }
Let us first define both the signal and background processes:
\begin{itemize}
	\item Signal: We concentrate in a final state that includes the semileptonic decay of the top quark, as previously mentioned, i.e., $e^-e^+\to tc\to \ell\nu_{\ell}bc$. Feynman diagram of the signal is shown in Fig. \ref{FD_eetc_LRSM}.
	\item Background: The main SM background processes come from $e^-e^+\to q_iq_j\ell\nu_{\ell}$, where $q_{i,\,j}$ are light quarks. These final states are originated mainly from $e^-e^+\to WW$ and $W$ breamsstrahlung $e^-e^+\to Wjj$. Besides, an irreducible SM background arises from $e^-e^+\to WW\to cb\ell\nu_{\ell}$ but it is negligible due to it depends on the Cabbibo-Kobayashi-Maskawa matrix element $V_{bc}$, which suppresses the process. The Feynman diagram of the dominant background is shown in Fig. \ref{BGDWW}.
\begin{figure}[!ht]
	\centering
	\includegraphics[scale=0.5]{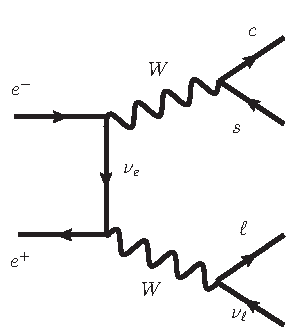}
	\caption{Feynman diagram of the dominant SM background.
	}
	\label{BGDWW}
\end{figure}	 
\end{itemize}

We begin our analysis by implementing the relevant Feynman rules via $\texttt{LanHEP}$ \cite{Semenov:2014rea} for $\texttt{MadGraph5}$ \cite{Alwall:2011uj}, later it is interfaced with $\texttt{Pythia8}$ \cite{Sjostrand:2008vc}. Afterward, the detector level simulation is performed with \texttt{Delphes 3} \cite{deFavereau:2013fsa} with the default CEPC card \cite{Chen:2017yel}. The kinematic distributions analysis was done via $\texttt{MadAnalysis}$ \cite{Conte:2012fm}, we used $\texttt{CT10}$ parton distribution functions \cite{Gao:2013xoa}, the jet finding package $\texttt{FastJet}$ \cite{Cacciari:2011ma} with the anti$-k_T$ algorithm \cite{Cacciari:2008gp} with a radius parameter $R=0.5$.

\begin{figure}[!ht]
	\centering
	\subfigure[]{	\includegraphics[width=0.4\textwidth]{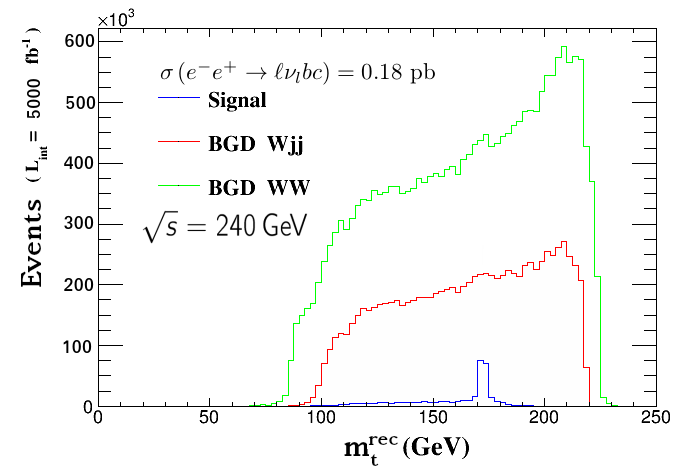}\label{TOPmassreconsLRSM}}
	\subfigure[]{		\includegraphics[width=0.4\textwidth]{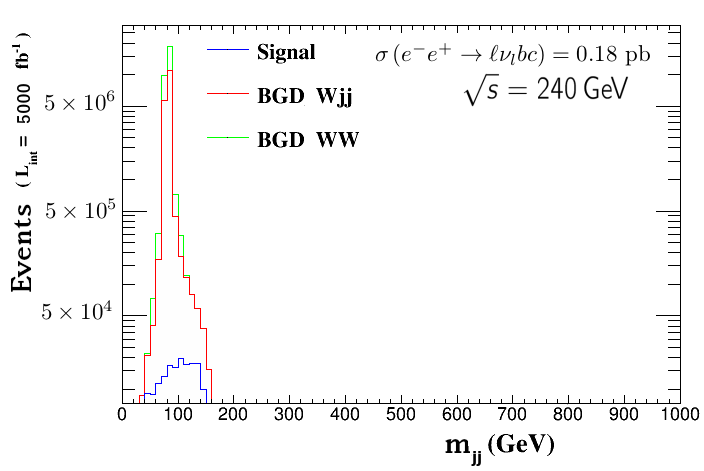}\label{WmassreconsLRSM}}
	\caption{Distributions for the signal and background. (a) Reconstructed top quark mass and (b) dijet invariant mass.
	}\label{DistKine}
\end{figure}

Fig. \ref{TOPmassreconsLRSM} shows the reconstructed mass of the top quark both for the signal and background processes setting the integrated luminosity to 5000 fb$^{-1}$, while in Fig. \ref{WmassreconsLRSM} we display the same but for the di-jet invariant mass distributions. Note that here we assume $\sigma(e^+e^-\to\ell \nu_{\ell} bc)=0.18$ pb in order to highlight the signal with respect to the background processes.

\begin{figure}[!ht]
	\centering
	\includegraphics[width=0.4\textwidth]{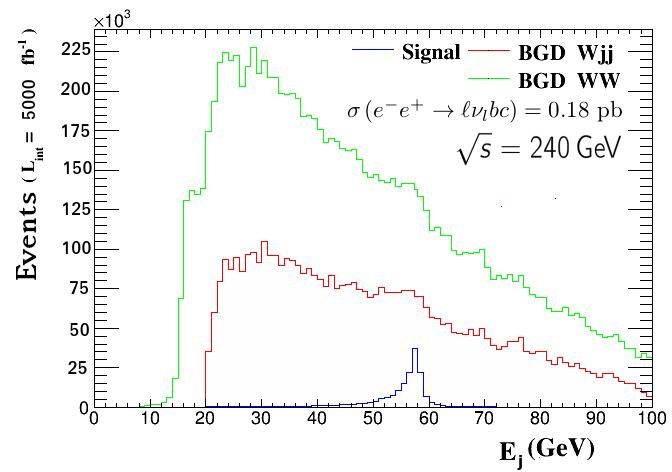} 
	\caption{Charm-jet energy distribution at $\sqrt{s}=240$ GeV for an integrated luminosity of 5000 fb$^{-1}$.
		\label{Ec}}
\end{figure}

A remarkable fact is a difference between the background and signal kinematic distributions. We notice in Fig. \ref{TOPmassreconsLRSM} the most important confirmation of the signal: the reconstruction of the top quark mass. Notwithstanding there is a neutrino in the final state and therefore it prevents a direct mass measurement of the top quark, it is possible to reconstruct $m_t$ from the knowledge of the center-of-mass energy and the charm jet energy as follows:
\begin{equation}
m_t^{\text{rec}}=\left(s-2\sqrt{s}E_c\right)^{1/2}.
\end{equation}
Moreover, we observe in Fig. \ref{WmassreconsLRSM} that the dijet invariant mass for the background events mainly reconstructs the $W$ gauge boson mass. Besides, because the signal involves two-body kinematics, the charm-jet energy is given by:
\begin{equation}
E_c=\left(\frac{\sqrt{s}}{2}-\frac{m_t^2}{2\sqrt{s}}     \right).
\end{equation}
For $\sqrt{s}=240$ GeV and $m_t=172.76$ GeV we have $E_c\sim 57.8$ GeV, as it is observed in Fig. \ref{Ec}.

\begin{figure}[!ht]
	\centering
	\includegraphics[width=0.4\textwidth]{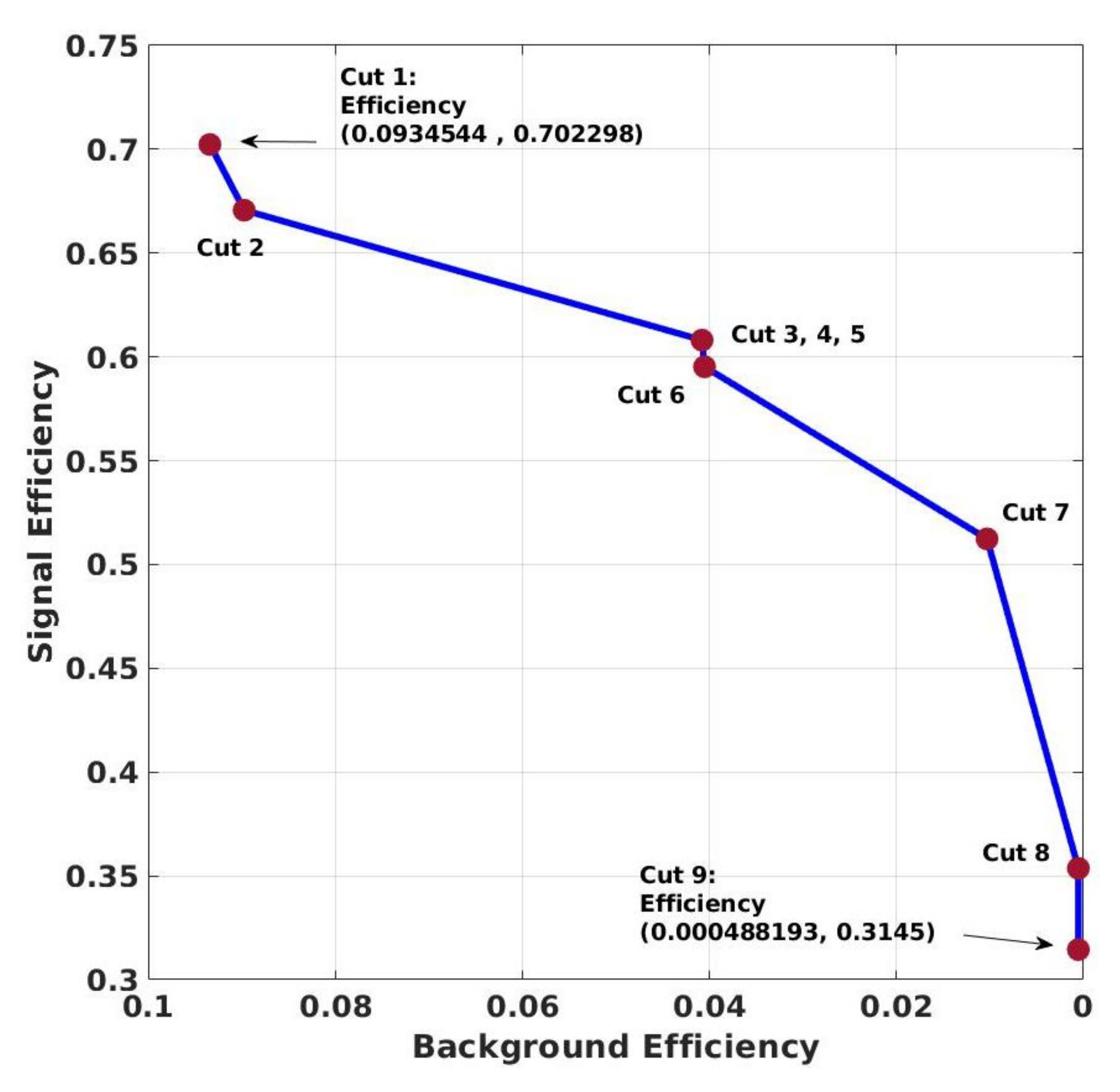} 
	\caption{Signal and background efficiencies after applying cuts. 
		\label{Eficiencia}}
\end{figure}

In order to isolate the signal from the background, we select events with exactly one charged lepton $\ell=e,\,\mu$ and at least two jets which one of them is tagged as a $b$-jet. We consider a $b$-tagging working point with 0.8 efficiency for the $b$-jet and a mistagging rate of 0.1 from $c$-jet \cite{Ruan:2018yrh}. In addition, motivated by the results shown in Figs. \ref{DistKine} and \ref{Ec}, we apply the following kinematical cuts:
\begin{enumerate}
	\item $E_{\ell}<60$ GeV,
	\item $p_T({\ell})>10$ GeV,
	\item $p_T(j)>20$ GeV,
	\item $|\eta_{\ell}|<3$,
	\item $|\eta_{j}|<3$,
	\item $|m_t^{\text{rec}}-m_t|<15$ GeV,
	\item $50<E_j$ GeV,
	\item $|m_{jj}-m_W|>10$ GeV,
	\item Due to the presence of a neutrino in the final state, we demand a missing energy greater than 20 GeV.
\end{enumerate}
Figure \ref{Eficiencia} shows the Signal-Background efficiencies plane after applying cuts. We observe that the most severe cuts are those associated with the reconstruction of the top quark mass, $m_{jj}$, and the energy of the c-jet.

Once the kinematical cuts were imposed, we found a substantial suppression on the background processes; then, we evaluate the signal significance, which is given by 
\begin{equation}
\sigma=\frac{\mathcal{N_S}}{\sqrt{\mathcal{\mathcal{N_S}+\mathcal{N_B}}}}
\end{equation}
where $\mathcal{N_S}$ and $\mathcal{N_B}$ are the signal and background events, respectively.

Our study focuses on the maximum energy to be achieved in CEPC, i.e., $\sqrt{s}=240$ GeV; however, motivated by the results of the production cross-section of the charm-top quark pair (see Fig. \ref{XS-energy}), we include the analysis for a center-of-mass energy of $\sqrt{s}=210$ GeV.

\begin{figure}[!ht]
	\centering
	\includegraphics[width=0.4\textwidth]{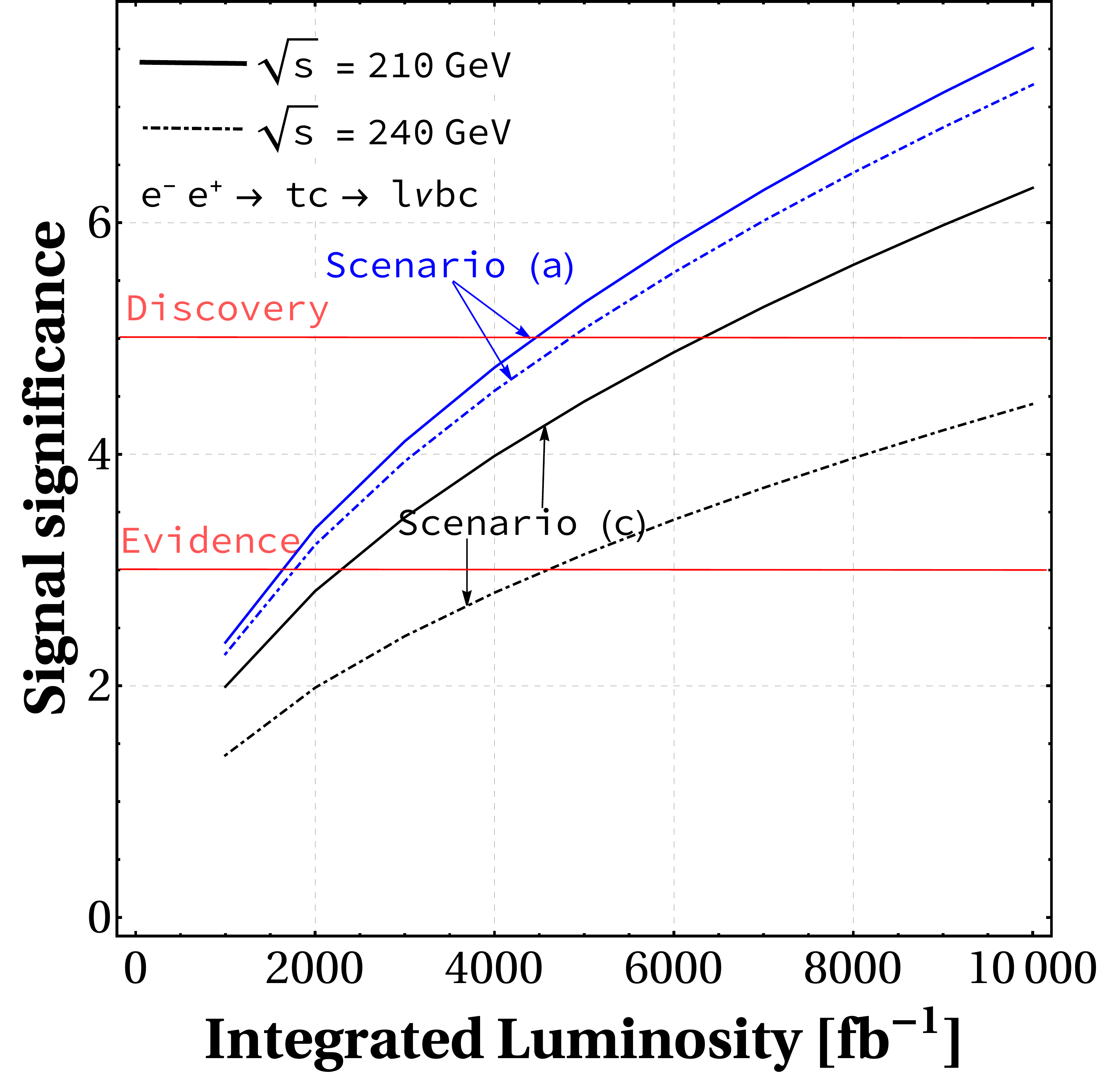} 
	\caption{Signal significance for the process $e^-e^+\to tc\to \ell\nu_{\ell}bc$ as a function of the integrated luminosity.
		\label{Significance}}
\end{figure}

Fig. \ref{Significance} shows the signal significance for $\sqrt{s}=210$ and $240$ GeV as a function of the integrated luminosity for scenarios $(a)$ and $(c)$. We omit scenario $(b)$ because it has a similar behavior as $(a)$. 

Fig. \ref{Significance} shows also our most important result in which we observe a potential signature for the process $e^-e^+\to V \to tc\to \ell_i \nu_i b c$ for $\sqrt{s}=240$ ($210$) GeV and integrated luminosities of about 4850 (4450) and 10000 (6320) fb$^{-1}$ for scenarios (a) and (c), respectively.  





\section{Conclusions\label{Conclusions}}

The FCNC processes are a key piece in providing a signal of models with new physics. In this work, we have studied FCNC process mediated by the vertex $Vtc$ involving the top and charm quarks, and a $V$ neutral gauge boson $(V=Z,\, Z^\prime)$. The $Z^\prime$ boson has arisen from the LRMM, which was our theoretical framework. The LRMM, as well as other extensions for the SM, increases the number of free parameters. However, we only focus on the parameters involved in the signal of our interests, $e^+e^- \to tc$.  To obtain the LHC and HL-LHC constraints values on the $Z^\prime$ mass, we analyze the production of  $Z^{\prime}$ boson via $pp\to Z^{\prime}$, with its subsequent decay to a pair of charged leptons $\ell\ell$. The allowed interval for the $Z^\prime$ mass is 4 TeV (4.6 TeV) $\lesssim m_{Z^\prime}$ for $\eta_{R}=0.1(1)$ using the LHC data, while the allowed mass interval is 5.8 TeV (6.4 TeV) $\lesssim m_{Z^\prime}$ for $\eta_{R}=0.1(1)$ if we consider the HL-LHC data.

As far as the mixing parameters are concerned, the reported values for Higgs boson signal strengths, $\mathcal{R}_{X}$ for $X=b, \tau , W, Z, \gamma$, were used to constrain $\eta_L$ as a function of $\cos\alpha$. The allowed region that satisfies all  Higgs boson signal strengths is $2\lesssim\eta_L\lesssim 2.2$ for $\cos\alpha=0.7$ as shown in Fig. \ref{calphaVSetaL}. Based on this allowed region, it has been possible to find the allowed region for $\lambda_R$ and $\Theta$ using the reported values for $g_V^u$ and $g_A^u$, as is shown in Fig. \ref{lamdRVSsinTheta}.  For instance, we can obtain $2.5 \lesssim\lambda_R\lesssim 4.7$ for  $\sin\Theta\approx 0.05$. The $\mathcal{BR}(t\to Z c)$ was also considered to constrain the mixing parameters, but in this case a correlation between $\left(\eta_L\right)_{tc}$ and $\left(\lambda_R\right)_{tc}$ was obtained for allowed and representative values of $\sin\Theta$. Fig. \ref{EtaL-LambR} shows us the behavior of the allowed regions for $\left(\eta_L\right)_{tc}$ and $\left(\lambda_R\right)_{tc}$ which is suppressed when $\sin\Theta$ increases. This behavior has been taken into account to propose a set of free parameters that are allowed. As a result of the analysis carried out in section 3, we have found a benchmark set with valid values, which are in agreement with the most up-to-date experimental constraints, for the free parameters involved in $e^+e^-\to V \to tc$, which are shown in Table \ref{ParamValues}. Based on these results, we proposed three scenarios, (a), (b), and (c), to search events for the process $e^+e^-\to V\to tc$. For the scenario (a), the signal coming from $e^+e^-\to V\to tc$ process could be detectable at $\sqrt{s}=240$ ($\sqrt{s}=210$) GeV with integrated luminosity close to 4850 (4450) fb$^{-1}$, even for masses of a new neutral gauge boson as high as  $m_{Z'}=7$ TeV. Meanwhile, for scenario (c) the signal will be detected if there is a higher integrated luminosity, of the order 10000 fb$^{-1}$ (6320) at $\sqrt{s}=240$ ($\sqrt{s}=210$). For scenario (b), the results show a behavior very close to scenario (a). The value of $\sqrt{s}=210$ GeV was suggested as a result analysis of $\sigma(e^+e^-\to tc)$, where it was noted that the contribution from the $Z$ gauge boson is the dominant. 

Thus, the CEPC, which was motivated to study in detail Higgs physics, can also be used to study top quark physics, taking into account the possibility to produce a single top quark in association with a charm quark using FCNC interactions.

\begin{acknowledgement}
Marco Antonio Arroyo Ure\~na especially thanks to \emph{PROGRAMA DE BECAS POSDOCTORALES DGAPA-UNAM} for postdoctoral funding. 
This work was supported by projects \emph{Pro\-gra\-ma de A\-po\-yo a Proyectos de Investigaci\'on e Innovaci\'on
	Tecnol\'ogica} (PAPIIT) with registration codes  IA106220 and IN115319 in \emph{Direcci\'on General de Asuntos de Personal
	Acad\'emico de Universidad Nacional Aut\'onoma de M\'exico} (DGAPA-UNAM), and \emph{Programa Interno de Apoyo para
	Proyectos de Investigaci\'on} (PIAPI) with registration code PIAPI2019 in FES-Cuautitl\'an UNAM and \emph{Sistema Nacional de Investigadores} (SNI) of the \emph{Consejo Nacional de Ciencia y Tecnolog\'ia} (CO\-NA\-CYT) in M\'exico. Tom\'as Antonio Valencia P\'erez is funded by Conacyt through the ‘Estancias posdoctorales nacionales’ program. We thankfully acknowledge computer resources, technical advise and support provided by Laboratorio Nacional de Superc\'omputo del Sureste de M\'exico.
\end{acknowledgement}


\end{document}